# The use of the logarithm of the variate in the calculation of differential entropy among certain related statistical distributions.

Author: Thomas M. Eccardt

Abstract:

*This paper demonstrates that basic statistics (mean, variance) of the logarithm of the variate itself can be used in the calculation of differential entropy among random variables known to be multiples and powers of a common underlying variate. For the same set of distributions, the variance of the differential self-information is shown also to be a function of statistics of the logarithmic variate. Then entropy and its "variance" can be estimated using only statistics of the logarithmic variate plus constants, without reference to the traditional parameters of the variate.*

Text:

Although the entropy or information content of a statistical distribution is an average of the logarithm of its probability density, the logarithm of the variate itself can occasionally play a role in the calculation of entropy. Likewise, the variance of logarithmic probability density, which we shall abbreviate here as "entropy variance," can be calculated through the use of the logarithmic variate. Jones (1979:137) has shown that if a random variable $\mathbf{X}$ has the probability density function $f(\mathrm{x})$, if the random variable $\mathbf{Y}$ is a function of that random variable, i.e. y = $g(\mathrm{x})$, and if H[$\mathbf{X}$] represents the entropy of $\mathbf{X}$, then

$$H[\mathbf{Y}] = H[\mathbf{X}] + \int f(\mathrm{x}) ln(g'(\mathrm{x})) dx, \qquad (1)$$

or

$$H[\mathbf{Y}] = H[\mathbf{X}] + E[ln(g'(\mathrm{x}))], \qquad (2)$$

if E[$\mathbf{X}$] is the expectation of the random variable $\mathbf{X}$.

This paper will utilize (2), along with an analogous formula for the variance of differential self-information, to show how entropy and entropy variance change with the multiples and powers of a given distribution. And since the average and variance of the logarithmic variates are functions of the multipliers and exponents used to create the distributions, formulas can be derived to calculate entropy and entropy variance based only the average and variance of the logarithmic variate. In particular, the formula for entropy involves only one constant, whose range appears to be quite narrow among common statistical distributions. And so there may be applications where the notion of an underlying distribution may be disregarded, and where entropy may be estimated directly from statistics of the logarithmic variate.

If the function of a random variable is $g(\mathrm{x}) = a\mathbf{X}^b$, then $g'(\mathrm{x}) = ab\mathbf{X}^{b-1}$, and

$ln(g'(x)) = ln(ab) + (b-1) \, ln(\mathbf{X})$.  So, by (2),

$$H[\mathbf{Y}] = H[\mathbf{X}] + E[ \, ln(ab) + (b-1) \, ln(\mathbf{X}) \, ]. \tag{3}$$

And since $E[a+b\mathbf{X}] = a + bE[\mathbf{X}]$ and $ln(ab) = ln(a) + ln(b)$,

$$H[a\mathbf{X}^b] = H[\mathbf{X}] + ln(b) + (b-1)E[ln(\mathbf{X})] + ln(a). \tag{4}$$

Clearly, then, the expectation of the logarithm of the variate itself, $E[ln(\mathbf{X})]$, can be used in the calculation of the differential entropy of a power of a random variable.  Of course, the variate must be restricted to values greater than zero, otherwise its logarithm and powers may be undefined or complex.  However if $\mathbf{X}$ has a minimum, then a constant can be added to the variate so as to make it always positive, without affecting its entropy.  Incidentally, when b = 1 equation (4) reduces to the well-known formula for the entropy of a multiple of a random variable (see, for example, Cover and Thomas (1991:233)),

$$H[a\mathbf{X}] = H[(\mathbf{X})] + ln(a). \tag{5}$$

The next step is to replace the multiplier and exponent with statistics of the logarithmic variate.  Clearly, the expectation is $E[ln(a\mathbf{X}^b)] = ln(a) + bE[ln(\mathbf{X})]$, and the variance is $V[ln(a\mathbf{X}^b)] = b^2V[ln(\mathbf{X})]$.  Solving these for $ln(a)$ and $b$ we get

$$ln(a) = E[ln(a\mathbf{X}^b)] - bE[ln(\mathbf{X})] \text{ and} \tag{6}$$

$$b = \sqrt{( \, V[ln(a\mathbf{X}^b)] / V[ln(\mathbf{X})] \, )}. \tag{7}$$

Equation (4) can be restated as the change in entropy due to the application of the exponent $b$ and the multiplier $a$ to the variate $\mathbf{X}$:

$$H[a\mathbf{X}^b] - H[\mathbf{X}] = (b-1)E[ln(\mathbf{X})] + ln(a) + ln(b). \tag{8}$$

Substituting (6) into (8) gives

$$H[a\mathbf{X}^b] - H[\mathbf{X}] = (b-1)E[ln(\mathbf{X})] + E[ln(a\mathbf{X}^b)] - bE[ln(\mathbf{X})] + ln(b), \tag{9}$$

or, with simplification,

$$H[a\mathbf{X}^b] - H[\mathbf{X}] = -E[ln(\mathbf{X})] + E[ln(a\mathbf{X}^b)] + ln(b). \tag{10}$$

Then if (7) is substituted into (10), the result is

$$H[a\mathbf{X}^b] - H[\mathbf{X}] = E[ln \, a\mathbf{X}^b] - E[ln(\mathbf{X})] \\ + ln(V[ln(a\mathbf{X}^b)]) / 2 - ln(V[ln(\mathbf{X})] )) / 2. \tag{11}$$

Clearly both sides of (11) represent a function as applied to $a\mathbf{X}^b$, less the same function applied to $\mathbf{X}$. This also proves that the entropy of any family of statistical distributions, $\mathbf{X}$, related by the transformation $a\mathbf{X}^b$, is equal to a function of the logarithmic variate, plus a constant, $K$, which varies with the type of underlying distribution:

$$H[\mathbf{X}] = E[ln(\mathbf{X})] + ln(V[ln(\mathbf{X})]) / 2 + K \qquad (12)$$

In some distributions (e.g. Weibull and lognormal) $a$ and $b$ are either parameters of the distribution or functions of the parameters. For those distributions, if the entropy of the distribution in general is known, and the mean and variance of the logarithmic variate are known (either through mathematical proof or through simulation), then $K$ can be calculated. So if a set of data is known to come from such a distribution, then the entropy can be calculated exclusively through the mean and variance of the logarithmic variate, without reference to its traditional parameters, which may be difficult to calculate. Entropy has often been difficult to estimate directly from statistical data, especially in continuous distributions, since the probability density of each value that the variate assumes must be reckoned. This usually results in downwardly biased estimates which increase as the size of the data set increases. Using only (a function of) the variate eliminates this problem. The fact that this new estimator uses only the logarithm of the variate should also reduce the bias which arises from data collection methods that tend to truncate or censor high-valued data points.

On the other hand, exact calculation of $K$ may not always be essential. Kapur (1989:68) has shown that when $E[ln(\mathbf{X})]$ and $E[ln(\mathbf{X})^2]$ are both prescribed, it is the lognormal distribution that has the highest entropy. This implies that the lognormal distribution has the highest $K$, at about 1.42. Though there may be no theoretical minimum for $K$, it appears that the most common positive distributions have a $K$ greater than or equal to unity. A range of 1-1.42 in nats rescales to a range of less than 2/3 bit. So if the required accuracy for an entropy estimate is a little less than one bit, then $K$ could be assumed to be equal to about 1.2, and the type of underlying distribution ignored. Another occasion where $K$ may be ignored would be when comparing the entropies of unknown or intractable distributions that can be shown to be related through the transformation $a\mathbf{X}^b$. Such distribution functions are easily recognized, having the same shape when graphed on a logarithmic scale. If only the relative entropies are sought, then the value of $K$ is irrelevant.

The generalized gamma (or Stacey) distribution provides a useful example to examine how equation (12) works in practice. This is due to the generalized gamma's self-reproductive property when multiplied or exponentiated. In other words, some of its parameters correspond to $a$ and $b$ in the transformed random variable, $a\mathbf{X}^b$. However, it has another parameter, $v$, whose values translate to the typical range of $K$, 1-1.42. Another advantage is that some of the commonest statistical distributions are special cases of the generalized gamma.

The probability density function of the generalized gamma distribution is as follows:

$GG[x \mid a,b,v] = b / (a^{bv} \Gamma(v)) \cdot x^{bv-1} \cdot exp\{-(x/a)^b\}$ (13)

A. Dadpay et al (2007:571) give its entropy as:

$H[\mathbf{X}_{GG}] = ln(a) + ln(\Gamma(v)) + v - ln(b) + (1/b - v)\Psi(v),$ (14)

where $\Gamma(v)$ is the gamma function of $v$, and $\Psi(v)$ is its digamma function, the differential of the logarithm of the gamma function. Using the logarithm of its Mellin transformation, two cumulants (mean and variance) of the logarithm of the generalized gamma distribution can be derived as:

$E[ln(\mathbf{X}_{GG})] = ln(a) + (1 / b)\Psi(v)$ (15)

and

$V[ln(\mathbf{X}_{GG})] = (1 / b^2)\Psi'(v),$ (16)

where $\Psi'(v)$ is the first derivative of the digamma function.

Substituting (15) and (16) into (12) gives

$H[\mathbf{X}_{GG}] = ln(a) + (1 / b)\Psi(v) + ln((1 / b^2)\Psi'(v)) / 2 + K.$ (17)

Setting (17) equal to (14) and solving for $K$ yields:

$K = ln(\Gamma(v)) - v\Psi(v) + v - ln((\Psi'(v)) / 2.$ (18)

The multiplier $a$ and the exponent $b$ are not present in (18), because $K$ is a constant for all powers and multiples of a given underlying distribution.

| $V$ | $-v\Psi(v)$ | $ln(\Gamma(v))$ | $-ln((\Psi'(v)) / 2$ | $K$ | Common distribution(s) |
|---|---|---|---|---|---|
| 30 | -101.5322 | 71.2570 | 1.69224 | 1.41704 | ≈ Lognormal |
| 10 | -22.5175 | 12.8018 | 1.12610 | 1.4104 | |
| 2 | -0.84556 | 0.0 | 0.21931 | 1.37375 | Chi Square |
| 1 | +0.57721 | 0.0 | -0.2488 | 1.32841 | Exponential, Weibull |
| ½ | +0.981755 | .572365 | -0.7981561 | 1.25596 | Half-normal, Chi |
| 0.125 | +1.0485 | 2.0184 | -2.0901 | 1.1018 | |
| 0.001 | +1.0006 | 8.9072 | -8.90775 | 1.00105 | |

*Table 1. Calculation of* **K** *for various forms of the generalized gamma distribution*

Each row of Table 1 lists a different form of the generalized gamma distribution based its $v$ parameter, the other corresponding terms of (18), and their total, $K$, plus some of the common names of the resulting distributions that are special cases of it. The first row, where $K$ equals 30, approximates the lognormal distribution; Lawless (1982:26) has demonstrated that the generalized gamma distribution approaches lognormality as $v$ approaches infinity.

When $v$ equals ½ (fifth row), the generalized gamma is equivalent to the standardized half-normal distribution, with the proviso that $b$ equals 2 and $a$ equals √2. Substituting these values into (15) and (16) yields E[$ln(\mathbf{X})$] + $ln$(V[$ln(\mathbf{X})$]) / 2 = -0.520185. Adding this to the value of $K$ = 1.25596 from Table 1 gives a total entropy of 0.735775, for the half-normal distribution with parameters μ = 0 and σ² = 1. Half of the support of the true standard normal distribution involves negative values, so its entropy cannot be calculated through equation (12). But because its probability density function is symmetrical about zero and because its absolute value can be generated by the half-normal distribution, it can be simulated by the half-normal, if only its sign is assigned by the tossing of a fair coin. The coin toss involves one bit of information, or 0.69315 nats, which, added to the half-normal entropy, 0.735775, gives about 1.42 for the standard normal, agreeing with Jones' (1979:138) derivation, ½$ln$(2πeσ²).

The last rows of Table 1 are supplied to show how $K$ approaches unity as $v$ approaches zero. Euler's infinite product formula for the gamma function clearly shows that as $v$ approaches zero, $\Gamma$(v) approaches 1/v. $\Psi$(v) also approaches 1/v, and $\Psi'$(v) approaches 1/v². So the third and fourth columns of Table 1 cancel, and $K$ approaches unity as $v$ approaches zero. Furthermore, it can be proven that $K$ equals unity for the Pareto and Uniform distributions.

It is gratifying to realize that several of the items in Table 1 are (limiting) distributions that result from combinations of independent distributions. The lognormal is the limiting distribution for the product of distributions. The gamma is the resulting distribution for the sum of gamma/exponential distributions, and the chi square results when the squares of normal distributions are added. And the Weibull results from repeatedly picking the minimum value of a fixed number of Weibull/exponential distributions. All of these distributions have a $K$ that ranges between 1.32841 and 1.41704. So if it can be assumed that a set of data results from a combination of independent factors as just described, then its entropy can be estimated with reasonable accuracy exclusively through statistics of the logarithmic variate, without any of the traditional parameters. In practice, if the data set contains negative values, the lowest negative value might be excluded, and its absolute value added to all other data points.

If it is surprising that entropy (or the mean of the self-information) of a random variable can be estimated through the mean and variance of the logarithmic variate itself, it is just as interesting, if not as useful, that the second central moment of the self-information can also be calculated using only the variance of the logarithmic variate. In sum, there is a surprisingly strong relation between the first two cumulants of self-information and the first two cumulants of the logarithm of the variate itself, at least among random variables that are multiples and powers of each other.

In order to investigate the second central moment of self-information, or the "entropy variance," we need to derive an equation analogous to (2). If $f$(x) represents a probability density function, then ordinary entropy, the first moment of self-information, can be represented as

H[X] = E[$ln$($f$(x))].                                                                (19)

As a second central moment, the entropy variance can be calculated similarly to the variance of any variate,

V[$ln$($f$(x))] = E[$ln^2$($f$(x))] − E$^2$[$ln$($f$(x))].                              (20)

Following Jones' (1979:137) treatment of entropy, we shall derive a formula for the entropy variance of a function, Y=$g$(X), of the random variable X.   We note in advance that the second term of the right side of (20) will simply be replaced by the square of the formula that Jones ultimately derives, namely,

H[**Y**]=H[**X**] + ∫$f$(x)$ln$($g'$(x))$dx$.                                          (21)

Now, the first term of (20) is equivalent to

E[$ln^2$($f$(y))] = ∫$f$(y) · $ln^2$($f$(y)) · $dy$.                                   (22)

We can follow Jones and substitute $f$(x)/$g'$(x) for $f$(y) and $dx$ · $g'$(x) for $dy$, to begin the derivation of a formula for a function of first term:

E[$ln^2$($f$(y))] = ∫$f$(x)/$g'$(x) · $ln$[$f$(x)/$g'$(x)] · $ln$[$f$(x)/$g'$(x)] · $dx$ · $g'$(x)          (23)

The two $g'$(x) factors cancel, and after the quotient law of logarithms is applied, restating the resulting equation as an expectation gives

E[$ln^2$($f$(y))] = E$^2$[ $ln$($f$(x)) - $ln$($g'$(x)) ].                             (24)

Carrying out the square, and re-writing the terms as separate expectations gives

E[$ln^2$($f$(y))] = E[ $ln^2$($f$(x) ] + E[ − 2 · $ln$($f$(x)) · $ln$($g'$(x)) ] + E[$ln^2$($g'$(x)) ],     (25)

which analogous to (2), only it represents the pure second moment (about zero) of the self-information for a function of the random variable **X**.

Rewriting (21) as expectations and squaring it yields

H$^2$[**Y**]=H$^2$[**X**] + 2 · H[**X**] · E[$ln$($g'$(x))] + E$^2$[$ln$($g'$(x))].             (26)

If we denote the entropy variance of the random variable Y as HV[Y], then clearly,

HV[**Y**] = E[$ln^2$($f$(y))] - H$^2$[**Y**].                                          (27)

Now (27) suggests that we subtract (26) from (25). But before we do, we can recognize that the parallel combinations of their first terms and then of their last terms separately represent (entropy) variances of their own:

$$HV[\mathbf{X}] = E[\ ln^2(f(x)\ ] - H^2[\mathbf{X}] \text{ and} \tag{28}$$

$$V[ln(g'(x))] = E[ln^2(g'(x))] - E^2[ln(g'(x))]. \tag{29}$$

So the resulting formula for the entropy variance of a function, Y, of the random variable X is

$$HV[\mathbf{Y}] = HV[\mathbf{X}] + V[ln(g'(x))] +$$
$$E[-2 \cdot ln(f(x)) \cdot ln(g'(x))\ ] - 2 \cdot H[\mathbf{X}] \cdot E[ln(g'(x))], \tag{30}$$

which is reminiscent of (2), though certainly more complicated. Again it is the original function plus or minus some other terms.

It will be convenient to treat the $ln(g'(x))$ terms of (30) separately when substituting $ln(g'(x)) = ln(ab) + (b-1)\ ln(\mathbf{X})$. The first term gives

$$V[ln(g'(x))] = V[ln(ab) + (b-1)\ ln(\mathbf{X})]. \tag{31}$$

Applying a well-known rule about the variance of the function of a random variable, namely, $V[a+b\mathbf{X}] = b^2 V[\mathbf{X}]$, gives

$$V[ln(g'(x))] = (b^2 - 2b + 1) \cdot V[ln(\mathbf{X})]. \tag{32}$$

Replacing $ln(g'(x))$ in the next term of (30) yields

$$E[-2 \cdot ln(f(x)) \cdot ln(g'(x))] = -2E[ln(f(x)) \cdot ln(ab) + ln(f(x)) \cdot (b-1) \cdot ln(\mathbf{X})], \tag{33}$$

or, after applying some familiar rules about expectations,

$$E[-2 \cdot ln(f(x)) \cdot ln(g'(x))] = -2E[ln(f(x)) \cdot ln(ab)] - 2(b-1)E[ln(f(x)) \cdot ln(\mathbf{X})]. \tag{34}$$

Since $H[\mathbf{X}] = -E[ln(f(x))]$, the last term of (30) can be rewritten as

$$- 2H[\mathbf{X}] \cdot E[ln(g'(x))] = 2 \cdot E[ln(f(x))] \cdot E[ln(g'(x))], \tag{35}$$

and when $ln(g'(x))$ is replaced it becomes

$$-2H[\mathbf{X}] \cdot E[ln(g'(x))] = 2E[ln(f(x)) \cdot ln(ab)] + 2(b-1)E[ln(f(x))] \cdot E[ln(\mathbf{X})]. \tag{36}$$

When (32), (34) and (36) are recombined, the $2E[ln(f(x)) \cdot ln(ab)]$ terms cancel, and grouping the terms into factors of either $b^2$ or $b$, we get

$$HV[\mathbf{Y}] = HV[\mathbf{X}] +$$
$$b^2 \cdot V[ln(\mathbf{X})] +$$
$$b \cdot (-2 \cdot V[ln(\mathbf{X})] + 2E[ln(f(\mathrm{x}))] \cdot E[ln(\mathbf{X})] - 2E[ln(f(x)) \cdot ln(\mathbf{X})])$$
$$+ V[ln(\mathbf{X})] - 2E[ln(f(x))] \cdot E[ln(\mathbf{X})] + 2E[ln(f(x)) \cdot ln(\mathbf{X})], \qquad (37)$$

which is the analog of (4).

This formula of entropy variance is again more complicated than that of ordinary entropy. However it is simpler in at least one way: the *a* variable is missing. This means that if one multiplies a random variable by a constant, its entropy variance is unaffected.

Since *b* is a function only of $V[ln(\mathbf{X})]$, i.e. $b = \sqrt{(V[ln(\mathbf{X}^b)] / V[ln(\mathbf{X})])}$, only the variance of the logarithmic variate will appear in the formula for the entropy variance of a multiple and power of a random variable:

$$HV[a\mathbf{X}^b] = HV[\mathbf{X}] +$$
$$V[ln(\mathbf{X}^b)] +$$
$$V^{\frac{1}{2}}[ln(\mathbf{X}^b)] \cdot V^{-\frac{1}{2}}[ln(\mathbf{X})] \cdot ( -2V[ln(\mathbf{X})] + 2E[ln(f(\mathrm{x}))] \cdot E[ln(\mathbf{X})] - 2E[ln(f(x)) \cdot ln(\mathbf{X})] )$$
$$+ V[ln(\mathbf{X})] - 2E[ln(f(x))] \cdot E[ln(\mathbf{X})] + 2E[ln(f(x)) \cdot ln(\mathbf{X})]. \qquad (38)$$

The result (38) is not nearly as simple as equation (12). The coefficients of $V[ln(\mathbf{X}^b)]$ and $V^{\frac{1}{2}}[ln(\mathbf{X}^b)]$ are complicated, and even include such unfamiliar statistics as $E[ln(f(x)) \cdot ln(\mathbf{X})]$, the average of the product of the logarithmic probability and the logarithmic variate, which would ordinarily be difficult to calculate. Unlike the simple constant coefficients of (12), the coefficients of (38) vary with the type of distribution, so there is no reasonable way to estimate entropy variance without regard to the distribution. Nevertheless, the coefficients are constant for multiples and powers of a given distribution, and the gist of (38) is that for any such distributions, entropy variance changes only with the variance and standard deviation of the logarithm of the variate itself.

But what is the meaning or use of entropy variance? If data from a given statistical distribution (continuous or discrete) are coded using an optimally efficient algorithm (with respect to code length), then the average length of the code is proportional to the entropy of the given data distribution. The variance of the code lengths will then be entropy variance of the distribution. Now if the encoded data are statistically independent from each other, then not only are the entropies additive, but also the entropy variances ought to be additive. For example, the average length of three consecutive code words is three times the length of one code word, and the variance of the length of three consecutive code words is three times that of one.

Although equations (4) and (37) apply strictly to continuous distributions, they represent approximations when applied to discrete distributions, if the probabilities are arranged by magnitude, and the variate is defined as the rank in the arrangement. For example, if the events of a space have probabilities 0.4, 0.2, 0.1 and 0.3, then the variates are assigned as 1, 3, 4, and 2, to rank the probabilities. Let us assume that the events are now sorted by probability, and that each is assigned an optimal code.  But now let a new code be created, by concatenating any one of, say, four equi-probable fixed-length prefixes to each code word, effectively quadrupling the cardinality of the probability space.   For each old code word, there will be four code words, and if the new code remains optimal, when sorted, all the words with different prefixes but the same ending (and therefore equal probability) should have neighboring rankings.  The rankings would have been equal, except for the need to disambiguate the equiprobable events.  On the other hand, all words with any given prefix will be about four ranks away from each other in the new code.  Now, since the new probability distribution is simply a subdividing (by 4) of the old distribution, the cumulative probability up to each use of any prefix in the distribution should match the cumulative probability of the code without that prefix.  Therefore the ranks (variates) of new code will be about 4 times higher than the ranks of the old code. In other words, the essential difference between the two distributions is that the variate of the new system is four times higher than that of the old.  Now the entropy of the new system will certainly be higher than that of the old system, since there are more events to code, and of course the code length is longer by the length of the prefix.  However the entropy variance of the rank distribution will be unchanged, since adding a constant to the random variable of the code length will not alter its variance.   As already mentioned, equation (37) tells us that multiplying the data in a continuous distribution by a constant will not affect its entropy variance.